\documentclass{aastex63}
\received{\today}
\submitjournal{ApJ}

\shorttitle{The Sun'€™s Giant Cellular Flows}
\shortauthors{Hathaway \& Upton}

\graphicspath{{./}}
\begin{document}

\title{Hydrodynamic Properties of the Sun'€™s Giant Cellular Flows}

\correspondingauthor{David H. Hathaway}
\email{dave.hathaway@comcast.net}

\author[0000-0003-1191-3748]{David H. Hathaway}
\affiliation{Stanford University, HEPL, Stanford, CA 94305, USA}

\author[0000-0003-0621-4803]{Lisa A. Upton}
\affiliation{Space Systems Research Corporation, Alexandria, VA, 22314}

\begin{abstract}

Measurements of the large cellular flows on the Sun were made by local correlation tracking of features (supergranules) seen in full-disk Doppler images obtained by the Helioseismic and Magnetic Imager (HMI) instrument on the NASA Solar Dynamics Observatory (SDO) satellite.
Several improvements made to the local correlation tracking method allowed for more precise measurements of these flows.
Measurements were made hourly over the nearly ten years of the mission-to-date.
A four-hour time lag between images was determined to give the best results as a compromise between increased feature displacement and decreased feature evolution.
The hourly measurements were averaged over the 34 days that it takes to observe all longitudes at all latitudes to produce daily maps of the latitudinal and longitudinal velocities.
Analyses of these flow maps reveal many interesting characteristics of these large cellular flows.
While flows at all latitudes are largely in the form of vortices with left-handed helicity in the north and right-handed helicity in the south, there are key distinctions between the low latitude and high latitudes cells.
The low latitude cells have roughly circular shapes, lifetimes of about one month, rotate nearly rigidly, do not drift in latitude, and do not exhibit any correlation between longitudinal and latitudinal flow.
The high latitude cells have long extensions that spiral inward toward the poles and can wrap nearly completely around the Sun.
They have lifetimes of several months, rotate differentially with latitude, drift poleward at speeds approaching 2 m s$^{-1}$, and have a strong correlation between prograde and equatorward flows.
Spherical harmonic spectral analyses of maps of the divergence and curl of the flows confirm that the flows are dominated by the curl component with RMS velocities of about 12 m s$^{-1}$ at wavenumber $\ell$ = 10.
Fourier transforms in time over 1024 daily records of the spherical harmonic spectra indicate two notable components - an $m = \pm\ell$ feature representing the low latitude component and an $m = \pm1$ feature representing the high latitude component.
The dispersion relation for the low latitude component is well represented by that derived for Rossby waves or r-modes.
The high latitude component has a constant temporal frequency for all $\ell$ indicating features advected by differential rotation at rates representative of the base of the convection zone high latitudes.
The poleward motions of these features further suggest that the high latitude meridional flow at the base of the convection zone is poleward - not equatorward.

\end{abstract}


\section{Introduction} \label{sec:intro}

The spectrum of the cellular convective motions at the surface of the Sun (and presumably at the surface other late-type stars) spans a broad range of cell sizes \citep{Hathaway_etal15}.
The smallest are granules with diameters of $\sim1000$ km, lifetimes of $\sim10$ minutes, and flow velocities $\sim4000$ m s$^{-1}$ (approaching the local speed of sound).
These convective cells are driven by radiative cooling at the Sun's photosphere \citep{SteinNordlund00, Nordlund_etal09}.
The spectrum of cellular flows extends all the way to the global scale (100s of Mm) with one notable feature - a broad but distinct bump representing supergranules.
Supergranules have  diameters of $\sim30$ Mm, lifetimes of $\sim18$ hours, and flow velocities of $\sim500$ m s$^{-1}$.
While there is no spectral evidence of any separation of scales associated with the giant cellular flows, there is a dynamical distinction.
As shown in \cite{Hathaway_etal15} the cellular flows with diameters greater than $\sim 200$ Mm (spherical harmonic degrees less than 20-30) are dominated by the effects of the Sun's rotation.
The flows are toroidal in the sense that they consist of vortices with much weaker diverging/converging flow components.
Here we examine the hydrodynamical properties of these giant cellular flows to ascertain their structures and dynamics, and to determine the roles they may play in producing the solar differential rotation, meridional circulation, and the magnetic dynamo responsible for the sunspot cycle.

The existence of giant cellular flows on the Sun was first proposed by \cite{SimonWeiss68} who coined the term ``giant cells'' for these structures.
They suggested that the solar convection zone should support cells that span the 200 Mm depth of the convection zone itself.
Early linear models of convection in rotating spherical shells (for both incompressible \citep{Gilman75} and compressible \citep{GlatzmaierGilman81} fluids) found that cells spanning the convection zone should be highly influenced by the Sun's rotation.
The cells tended to align with the rotation axis (the Taylor-Proudman Theorem) and to stretch north/south across the equator.
The Coriolis force acting on the flows in these ``banana'' cells produces an equatorward flux of angular momentum (a Reynolds stress) that would drive/maintain solar differential rotation with a rapidly rotation equator.

Over the intervening years, numerical models allowed for highly nonlinear flows and encompassed more and more density scale heights so as to bring the modeled domain closer to spanning the entire convection zone \citep{Gilman79, Miesch_etal00, BrunToomre02, Miesch_etal08, Hotta_etal15, Nelson_etal18}.
As the modeled flows became more turbulent, the simple banana cell structure became less apparent, particularly in the near surface layers.
Nonetheless, these models showed that the cellular flows produced Reynolds stresses which were critically important for maintaining the Sun's differential rotation.
In these simulations, the cells propagate prograde at low latitudes and retrograde at high latitudes.
The cellular flows themselves have negative kinetic helicity (the dot product of velocity and the curl of velocity) in the north and positive kinetic helicity in the south throughout the bulk of the convection zone.
(Note that in the axisymmetric mean field models of solar/stellar convection \citep{DurneySpruit79, Hathaway84, KichatinovRudiger93, KichatinovRudiger05}, the Reynolds stresses are parameterized  and referred to as the ``$\Lambda$-effect''.)

Observational evidence for giant cells was slow in coming.
\cite{Bumba70} noted the existence of large magnetic structures that might be associated with giant cells but we now know that these structures are a consequence of the transport of magnetic flux by differential rotation, meridional flow, and supergranule diffusion \citep{DeVore_etal85}.
The advent of continuous, full-disk observations of the Sun in the mid-1990s provided new data appropriate for observing and characterizing giant cells.
\cite{Hathaway_etal96} used direct Doppler observations from the Global Oscillations Network Group (GONG) network of instruments and found that large-scale Doppler features were evidenced by long-lived features rotating at the Carrington rotation rate.
\cite{Beck_etal98} used direct Doppler observations from the Michelson Doppler Investigation (MDI) instriment on the ESA/NASA Solar and Heliospheric Observatory (SOHO) mission \citep{Scherrer_etal95} and also found long-lived features that rotated at the Carrington rate.
They noted that the low latitude features they found were extended in longitude but with narrow latitudinal extents - unlike the banana cells in theoretical models.
\cite{Lisle_etal04} also used direct Doppler observations from MDI and found that supergranules tend to align with each other in a north-south direction over a wide range of latitudes - consistent with the banana cells in theoretical models.

\cite{Hathaway_etal13} tracked the motions of groups of supergranules seen in Doppler data from the Helioseismic and Magnetic Imager (HMI) instrument on the NASA Solar Dynamics Observatory (SDO) satellite \citep{Scherrer_etal12} and produced the first images of giant cellular flows.
The hydrodynamic properties of these flows included kinetic helicity and Reynolds stresses like those in the simulations, but the cellular structure was quite different - dominated by long, narrow cells at high latitudes with peak flow velocities of $\sim20$ m s$^{-1}$.
Shortly thereafter, \cite{Bogart_etal15} used the helioseismic technique of ring-diagram analysis and found nearly identical structures that coincided in space and time with those found by \cite{Hathaway_etal13}, but with peak flow velocities of only $\sim0.5$ m s$^{-1}$.

Recently, \cite{Loptien_etal18} tracked the motions of granules seen in intensity images from SDO/HMI and found giant cellular structures in the equatorial region that had properties associated with Rossby waves \citep{Rossby39} or r-modes \citep{Saio82, WolffBlizard86} - waves in which horizontal pressure gradients are balanced by the Coriolis force on the associated flows.
Observations were limited to the lower latitudes but indicated flow velocities of $\sim10$ m s$^{-1}$ at $\pm50^\circ$ latitude.
A ring-diagram analysis by \cite{Proxauf_etal20} has provided depth information indicating a slight decrease in amplitude with depth down to 8 Mm.

Here we improve and expand on the observations of \cite{Hathaway_etal13}.
We track the motions of supergranules seen in Doppler observations from SDO/HMI.
The Doppler signal associated with these largely horizontal flows becomes strongest near the observed limb of the Sun.
This characteristic allows for flow measurements at high latitudes - latitudes largely unexplored by other methods (helioseismology, tracking granules in intensity images). 
We find that the giant cellular flows on the Sun have two distinct regimes - a high-latitude regime with long, narrow cells that spiral into the polar regions, and a low-latitude regime with more circular cells that propagate like Rossby waves or r-modes.
The high latitude cells have hydrodynamic properties that impact the global dynamics of the Sun's convection zone and the mechanisms associated with its magnetic dynamo.

\section{Data and Data Preparation} \label{sec:data}

The data we used to measure the giant cell flows are full-disk Dopplergrams obtained by SDO/HMI.
A set of individual filtergrams at 24 wavelengths/polarizations were obtained every 45 s, each was registered to a central time, and then averaged over 720 s with a tapered temporal filter by the instrument team to produce Doppler images largely free of signal from the 5-minute p-mode oscillations.
We took these 720 s Dopplergrams and found and removed the signals fixed relative to the visible disk (spacecraft motion, solar rotation/differential rotation, meridional flow, and convective blue shift) see \cite{Hathaway_etal15} for further details concerning this process.

The resulting Dopplergrams are dominated by the signals from solar supergranules.
Since these flows are largely horizontal, the Doppler signal peaks near the limb.
This gives good signal at high latitudes (unlike the radial flow signal associated with the p-modes used in helioseismology).

These $4096^2$ full-disk Doppler images of the solar supergranulation were mapped to $4096^2$ pixels in equirectangular heliographic coordinates (4096 equispaced positions in longitude from $90^\circ$ east of the central meridian to $90^\circ$ west of the central meridian and 4096 equispaced position in latitude from the south pole to the north pole).
The Doppler signals at targeted longitude/latitude positions were determined using bi-cubic interpolation from the $4\times4$ pixels surrounding the location in the full-disk image.

These $4096^2$ heliographic maps of the supergranule Doppler signal were convolved with a Gaussian blurring function and then resampled to produce $512^2$ maps for the local correlation tracking procedure.
This provided sufficient spatial resolution to resolve the supergranules while minimizing the computational effort in tracking these features.

\section{Local Correlation Tracking Procedure} \label{sec:LCT}

We measured the proper motions of supergranules using a local correlation tracking (LCT) method \citep{November_etal87} like that used in \cite{Hathaway_etal13}.
The cross-correlation coefficient, $\chi$, is a number between $\pm 1$ given by

\begin{eqnarray}\label{eqn:XC}
\chi(\theta,\phi,\Delta\theta,\Delta\phi,\Delta t) = & \int \int [V(\theta,\phi,t) - \overline {V(\theta,\phi,t)}] [V(\theta + \Delta\theta ,\phi + \Delta\phi,t + \Delta t) -   \overline {V(\theta + \Delta\theta ,\phi + \Delta\phi,t + \Delta t)}] d\theta d\phi \nonumber \\ &
/ [\sigma(\theta,\phi,t) \sigma(\theta + \Delta\theta ,\phi + \Delta\phi,t + \Delta t)]
\end{eqnarray}

\noindent
where $\sigma$ is the RMS variation given by

\begin{equation}\label{eqn:sigma}
\sigma^2(\theta,\phi,t) = \int \int [V(\theta,\phi,t) - \overline {V(\theta,\phi,t)}]^2 d\theta d\phi 
\end{equation}

\noindent
with $V(\theta,\phi,t)$ being the Doppler velocity at co-latitude $\theta$, longitude $\phi$ and time $t$.
The overbar indicates an average over the area within the correlation window while $\Delta t$ is the time difference and $\Delta\theta$ and $\Delta\phi$ are the spatial offsets between the observations.

The horizontal velocity vector at a given location was determined by the displacement giving the highest correlation for a patch of data within the correlation window in one map when correlated to a similar patch in a second map obtained at a predetermined time lag (e.g. 4-hr, 8-hr, or 16-hr).
We made these determinations at a $256^2$ array of points equispaced on the Doppler velocity maps.
While the LCT method closely follows that used in \cite{Hathaway_etal13}, several improvements have been made based on a recent study \citep{Mahajan_etal20}.

\cite{Hathaway_etal13} used a circular correlation window with a 21-pixel diameter ($\sim 90$ Mm at the equator).
Here we used an oval window that becomes elongated in longitude with latitude by a factor of $\csc \theta$ so as to cover the same physical area on the surface of the Sun at all latitudes (up to $75^\circ$ latitude, beyond which the oval remains 21-by-81 pixels in size).
We also expanded the search area (larger maximum $\Delta\theta$ and $\Delta\phi$) in our search for the best correlation.
This assures a full sampling of possible displacements.

We modified the method for determining the location of the correlation maximum.
We find an estimate for the location of the peak to a fraction of a pixel by calculating the locations of the peaks of the parabolas passing through the maximum correlation in both the latitudinal and longitudinal directions.
This estimate is used to then shift (using a bi-cubic interpolator) the original patch of data by that fraction of a pixel to then repeat the process and better determine the fractional pixel shift.

A final revision to the LCT procedure was to use a 4-hr time-lag along with the 8-hr and 16-hr lime-lags used in \cite{Hathaway_etal13}.
The different time-lags have important consequences.
The shorter the time-lag the smaller the displacement and the larger the error in the measurement.
The longer the time-lag the more the evolution of the supergranules within the correlation window and the larger the error in the measurement.
Experiments with time-lags from 1-hr to 24-hr indicate, for this data, a minimum in measurement noise at a time-lag of 4-hr - the nominal time-lag used for the bulk of this study.
Another consequence of using different time-lags is sensitivity to flows at different depths within the Sun.
\cite{Hathaway12A} showed that tracking supergranules with longer time-lags gives results that are sensitive to longer-lived supergranules that tend to be larger and to extend to deeper depths.
In fact, supergranules were found to be advected by flows at depths equal to their diameters and proportional to the time-lag used.
Thus, we can use the different time-lags to obtain information about changes in the flow structure with depth.

\begin{figure}[htb]
% \centering
\includegraphics[width=1.0\columnwidth]{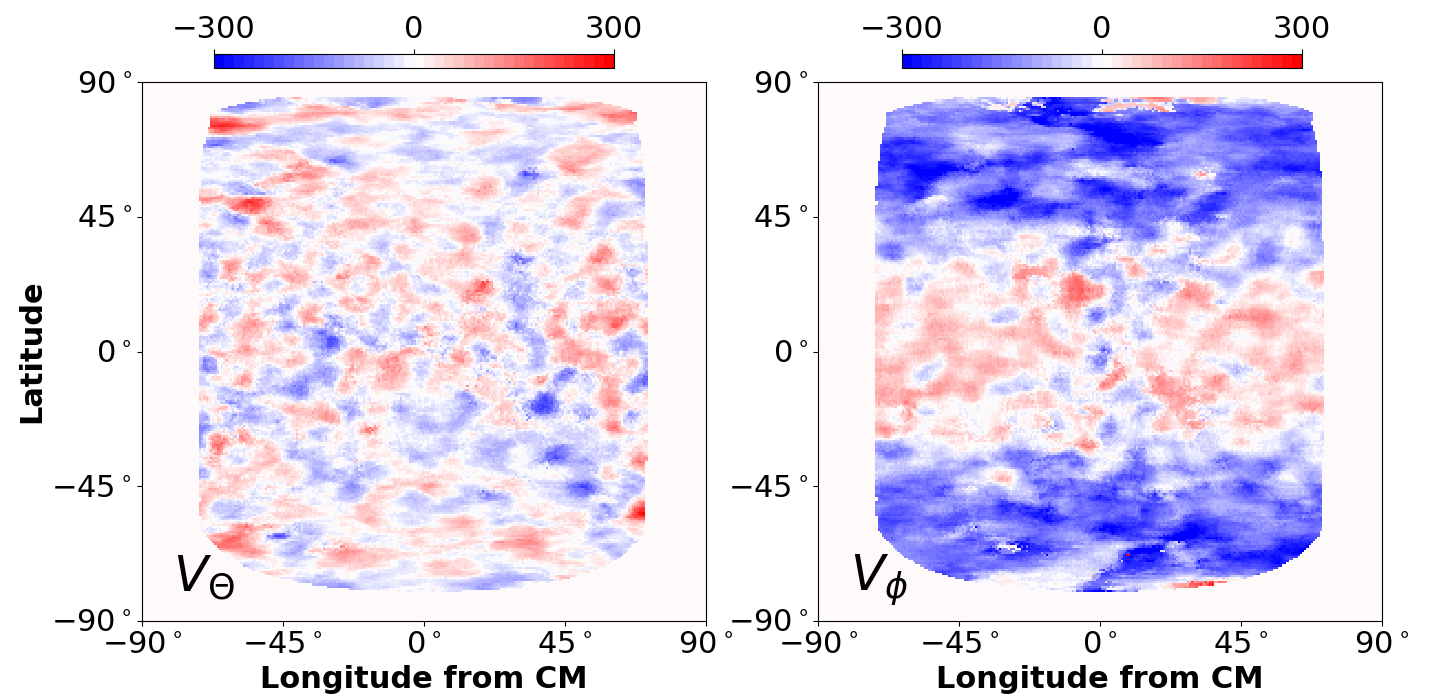}
\caption{Latitudinal (left panel, northward in red, southward in blue) and longitudinal (right panel, prograde in red, retrograde in blue) velocities determined from local correlation tracking using a single pair of Doppler maps separated by 4-hr in time. Both have a velocity scale of $\pm300$ m s$^{-1}$.}
\label{fig:LCTsample}
\end{figure}

An example of a raw LCT flow map obtained with the 4hr time-lag is shown in Figure~\ref{fig:LCTsample}.
The latitudinal velocity, $V_\theta$, is shown in the left panel.
The longitudinal velocity, $V_\phi$, is shown in the right panel.
Both have a velocity scale of $\pm300$ m s$^{-1}$.
The longitudinal velocity is dominated by the differential rotation of the supergranule pattern with large retrograde flow at high latitudes and prograde flow near the equator.
(Note that all velocities are relative to the frame of reference rotating with the Carrington period of 27.2753 days synodic.)
The latitudinal velocity shows a hint of the $\sim15$ m s$^{-1}$ poleward meridional flow.
Both velocity components exhibit a mottled pattern associated with the large cellular flows.
Similar maps are obtained with the 8hr and 16hr time-lags but with higher noise levels, with different differential rotation and meridional flow velocities, and slight differences in the mottled patterns.
These raw 4-hr LCT flow maps were constructed at hourly intervals over the length-to-date of the SDO Mission (2010 May through 2020 March).
Flow maps at 8-hr and 16-hr were obtained over more limited intervals of the mission.

\section{Giant Cell Flow Map Construction Procedure} \label{sec:GiantCellMaps}

The raw LCT flow maps described in the previous section were averaged together to produce giant cell flow maps.
(Typically some 800 raw hourly LCT flow maps obtained over 34 days are averaged together to make each of the daily giant cell flow maps.)
Here, again, we improve upon the method used in \cite{Hathaway_etal13}.

\begin{figure}[htb]
% \centering
\includegraphics[width=1.0\columnwidth]{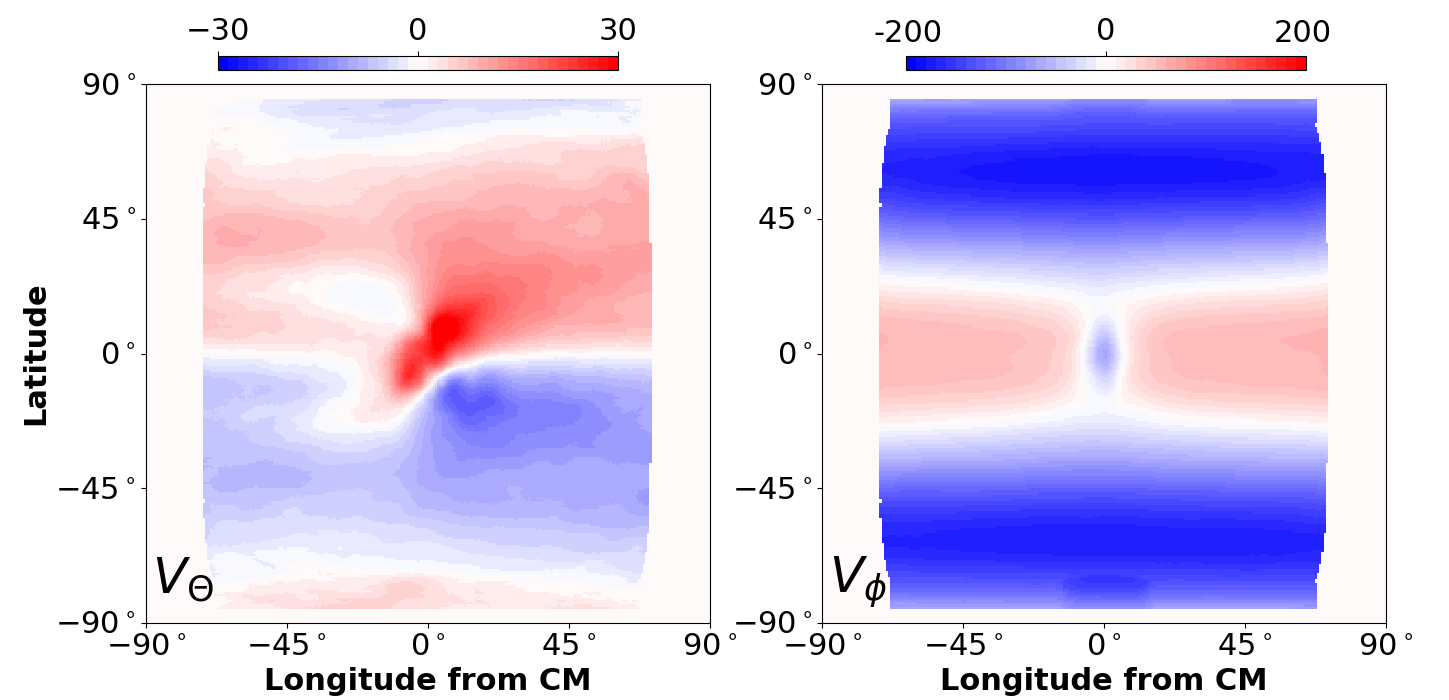}
\caption{Disk averaged flows. Latitudinal (left panel, northward in red, southward in blue, $\pm30$ m s$^{-1}$) and longitudinal (right panel, prograde in red, retrograde in blue, $\pm200$ m s$^{-1}$) velocity signals when averaged over all raw LCT flow maps at the 4-hr time-lag.}
\label{fig:DiskAveragedFlows}
\end{figure}

\cite{Hathaway_etal13} determined the differential rotation and meridional flow as functions of latitude only, and then removed those axisymmetic flow signals from the raw LCT flow maps before averaging.
We have found that the differential rotation and meridional flow signals have components that appear to vary in longitude.
This is illustrated in Figure~\ref{fig:DiskAveragedFlows} which shows the signal averaged over all raw LCT flow maps at the 4-hr time-lag.
This long average removes the mottled signal due to the large cellular flows and reveals the signals due to meridional flow (left) and differential rotation (right).
Both flow components exhibit some apparent variation with longitude relative to the central meridian.
We attribute this spurious signal component to line-of-sight projection effects on the supergranule Doppler signal (cf. \cite{Hathaway_etal06}).
It is this disk averaged signal, with its added longitudinal variation, that is removed from the raw LCT flow maps to better reveal the giant cell flows.

After removing these disk averaged signals associated with the axisymmetric flows, the raw LCT flows maps are averaged together using the Carrington longitude of the central meridian associated with each map as a reference for positioning the data.
\cite{Hathaway_etal13} constructed giant cell flows maps in the traditional manner for individual Carrington rotations - using the Carrington longitude for each raw LCT flow map as the longitude to use at all latitudes.
This commonly used method neglects the Sun's differential rotation.
The method results in features near the equator being pinched together and features near the poles being stretched apart relative to where they actually are on the Sun at a given time.
More than $360^\circ$ in longitude on the Sun are placed on these maps at the equator due to a rotation period three days shorter than the Carrington period.
As little as $285^\circ$ in longitude on the Sun are stretched out near the poles due to a rotation period seven days longer than the Carrington period.

Here we average the raw LCT flow maps using longitudes based on differential rotation.
The flow data at each latitude is shifted in longitude to its place on a giant cell flow map based on a measured differential rotation profile.
This process allows us to construct giant cell flow maps of the entire solar surface for any given moment in time (assuming evolution of the flow pattern on a timescale longer than the rotation period).
The process does, however, requires us to measure the differential rotation profile of the giant cell flow pattern itself.

\begin{figure}[htb]
% \centering
\includegraphics[width=1.0\columnwidth]{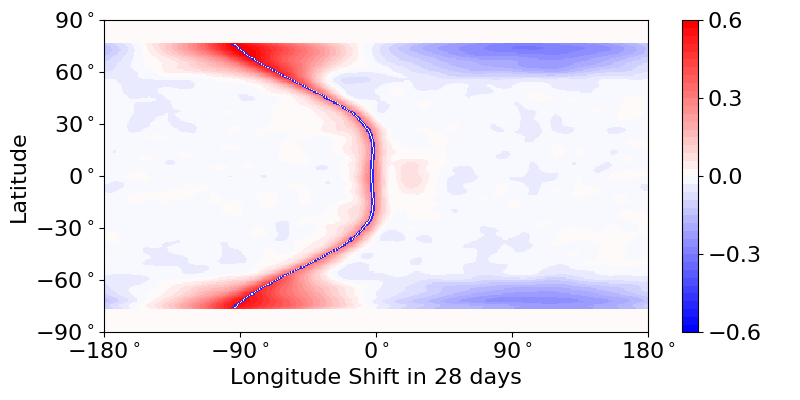}
\caption{Cross-correlation amplitude (red positive and blue negative) for a 28-day time-lag between temporary giant cell flow maps as a function of longitudinal shift and latitude. The three pixels surrounding the correlation peak at each latitude are colored blue. A polynomial fit to the peak locations (Eq.~\ref{eqn:GCDR1}) is indicated by a white line through the blue pixels.}
\label{fig:Xcor28Days}
\end{figure}

We find the differential rotation profile for the giant cell features by cross-correlating longitudinal strips of velocity data from temporary giant cell flow maps constructed 28 days apart.
(The temporary giant cell flow maps are constructed in the traditional manner used in \cite{Hathaway_etal13} but are made on a daily basis instead of a Carrington rotation period basis.)
We cross-correlate data for the full range of longitude shifts and average together all the cross-correlations.
The results are shown in Figure~\ref{fig:Xcor28Days}.
A functional fit through the locations of the cross-correlation peak at each latitude gives the longitude shift, $\delta\phi$, for the giant cellular features as

\begin{equation}\label{eqn:GCDR1}
\delta\phi(\theta,\Delta t) =  (-3^\circ + 28^\circ \cos^2\theta  - 220^\circ \cos^4\theta 
+ 90^\circ \cos^6\theta) \Delta t/28^d
\end{equation}

\noindent
where $\Delta t$ is the time difference (relative to the target time for the giant cell flow map) of the raw LCT flow map being added to the average.
(Aspects of this differential rotation profile, and the associated meridional flow profile will be discussed in Section~\ref{sec:DRMF}.)

\begin{figure}[htb]
\includegraphics[width=1.0\columnwidth]{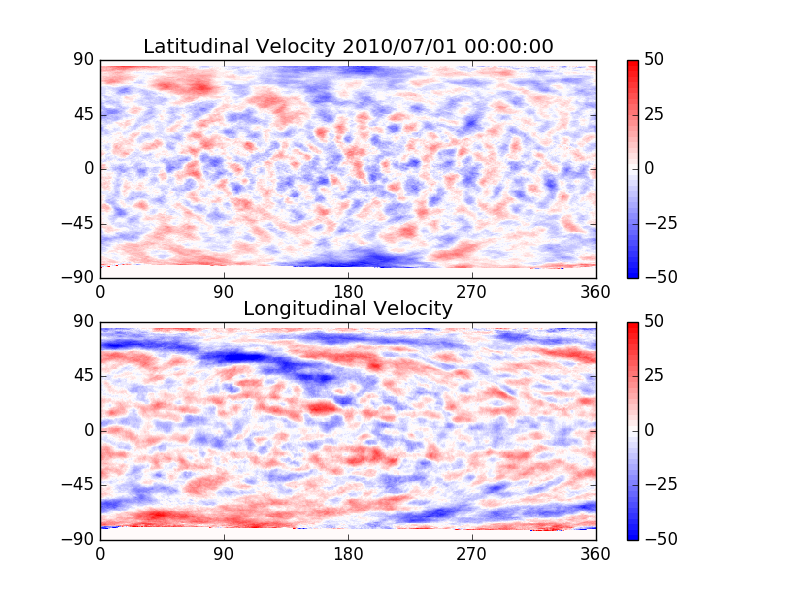}
\caption{Giant cell flow map example from $00^h:00^m:00^s$ UT on 2010 July 1. The upper panel shows the latitudinal velocity (red northward). The lower panel shows the longitudinal velocity (red prograde).}
\label{fig:GCFlowMaps}
\end{figure}

We use this longitude shift (Eq. \ref{eqn:GCDR1}) to construct maps of the giant cell flows on a daily basis.
Given the target day and time, we average some 800 raw LCT flow maps from $\pm 17d$ so as to include all longitudes even at the highest latitudes.
The velocity data at each latitude is shifted in longitude according to Eq.~\ref{eqn:GCDR1} to where it would have been at the target time.
An example of one of these daily giant cell flow maps is shown in Figure \ref{fig:GCFlowMaps}.

\section{Giant Cell Morphologies and Lifetimes} \label{sec:Morphology}

Inspecting the 3000+ daily giant cell maps over the nearly 10 years of the SDO mission-to-date reveals key morphological characteristics of the cellular flows.
The low latitude cells are nearly circular in shape while the high latitude cells are elongated in longitude to form a spiral pattern centered on the poles.

\begin{figure}[htb]
\includegraphics[width=1.0\columnwidth]{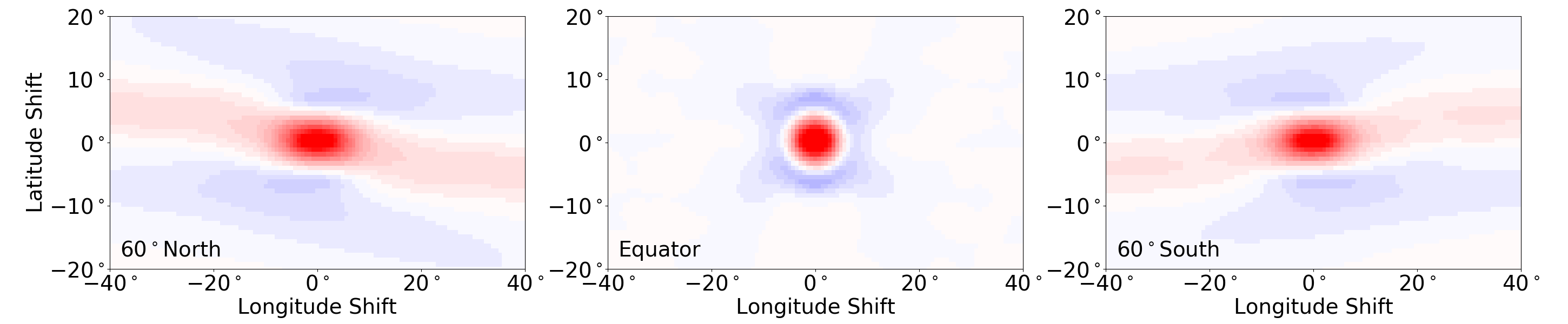}
\caption{Auto-correlation maps (red positive, blue negative correlation) of the vorticity at three latitudes - $60^\circ$ North (left panel), the equator (middle panel), and  $60^\circ$ South (right panel). Cells at high latitudes are sheared out in longitude to form long narrow tilted cells. Cells at the equator are nearly circular.}
\label{fig:GCAutoCor}
\end{figure}

This is born out by auto-correlating strips of data at different latitudes as shown in Figure~\ref{fig:GCAutoCor}.
Vorticity maps were constructed from each set of daily giant cell flow maps (Figure~\ref{fig:GCFlowMaps}) by taking the curl of the horizontal flow vectors.
Three fairly broad ($\pm 22.5^\circ$) strips (the full $360^\circ$ in longitude) from the vorticity maps, centered on the equator and at $\pm 60^\circ$, were auto-correlated with themselves to reveal the cell morphology at those locations.
Auto-correlation maps were constructed by shifting each vorticity map strip through $\pm 180^\circ$ in longitude and $\pm 22.5^\circ$ in latitude.

The average auto-correlation maps in Figure~\ref{fig:GCAutoCor} show explicitly that the cells at high latitudes are long and narrow with characteristic tilts that produce a spiral pattern around the poles.
The low latitude cells are nearly circular with no evidence of any tendency to be elongated north to south or east to west.

The lifetimes of the cells can be estimated from the strength of the cross-correlation shown in Figure~\ref{fig:Xcor28Days}.
Similar cross-correlations were calculated for the longer time-lag of 56 days.
The cross-correlation at the equator drops from its initial 1.0 to 0.3 after 28 days and to 0.07 after 56 days - indicating lifetimes between one and two months for the cells near the equator.
The cross-correlation at latitudes of $\pm 60^\circ$ drops from its initial 1.0 to 0.5 after 28 days and to 0.3 after 56 days - indicating lifetimes of several months for the high latitude cells.

This difference in both morphology and lifetimes of the cells at low vs. high latitudes is the first indication of two different dynamical regimes for these large cellular flows.

\section{Giant Cell Differential Rotation and Meridional Motion} \label{sec:DRMF}

We measure the differential rotation and meridional motion of the giant cells by finding the displacement in longitude and latitude that gives the maximum correlation for narrow longitude strips from giant cell flow maps separated by 28 days in time.
These measurements were made on six month long intervals of data so as to obtain estimates of uncertainty and variability.

\begin{figure}[htb]
\includegraphics[width=1.0\columnwidth]{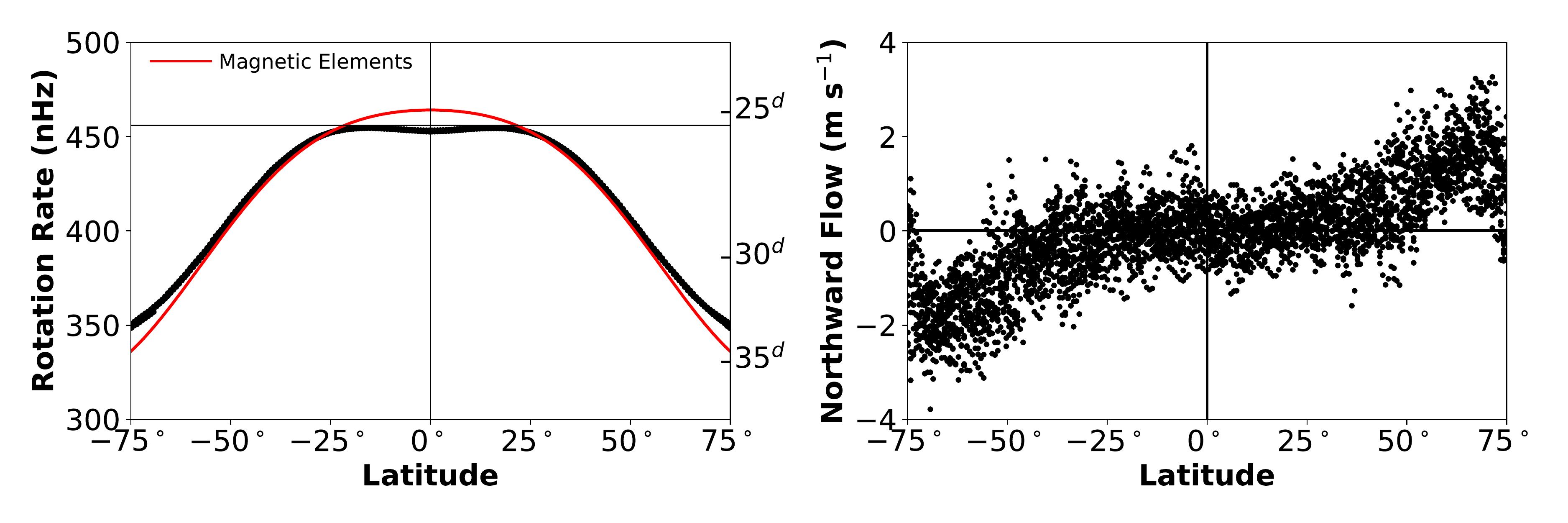}
\caption{left Panel: Sidereal differential rotation rate of the giant cells measured over 19 6-month intervals. The rotation rate derived from the motions of the small magnetic elements by \cite{HathawayRightmire11} is shown in red for reference. The horizontal line is at the Carrington rate of 456 nHz. Corresponding sidereal rotation periods are indicated on the right-hand axis. Right Panel: Meridional motion of the giant cells measured over 19 6-month intervals. There is little or no meridional motion of the cells between $\pm25^\circ$. At higher latitudes the meridional motion is poleward at velocities approaching 2 m s$^{-1}$.}
\label{fig:GCDRMF}
\end{figure}

The differential rotation profiles for each of these intervals are shown in the left panel of Figure~\ref{fig:GCDRMF} along with, for reference, the rotation rate of the small magnetic elements as given by \cite{HathawayRightmire11}.
A surprising feature of this profile is the relatively flat section within about $25^\circ$ of the equator with rotation slightly slower than the Carrington rate.
Both the flatness and the slowness are not exhibited by other solar features at these latitudes at any depth within the Sun.

The meridional motion profiles for each of the six-month intervals are shown in the right panel of Figure~\ref{fig:GCDRMF}.
Here again we find a flat section within about $25^\circ$ of the equator.
The meridional motion of the giant cell pattern at higher latitudes is poleward in each hemisphere with a peak velocity approaching 2 m s$^{-1}$.

These axisymmetric flow measurements provide another indication of the two different dynamical regimes - one at low latitudes and another with different characteristics at high latitudes.

\section{Giant Cell Kinetic Helicity and Reynolds Stress} \label{sec:KHRS}

\begin{figure}[htb]
\includegraphics[width=1.0\columnwidth]{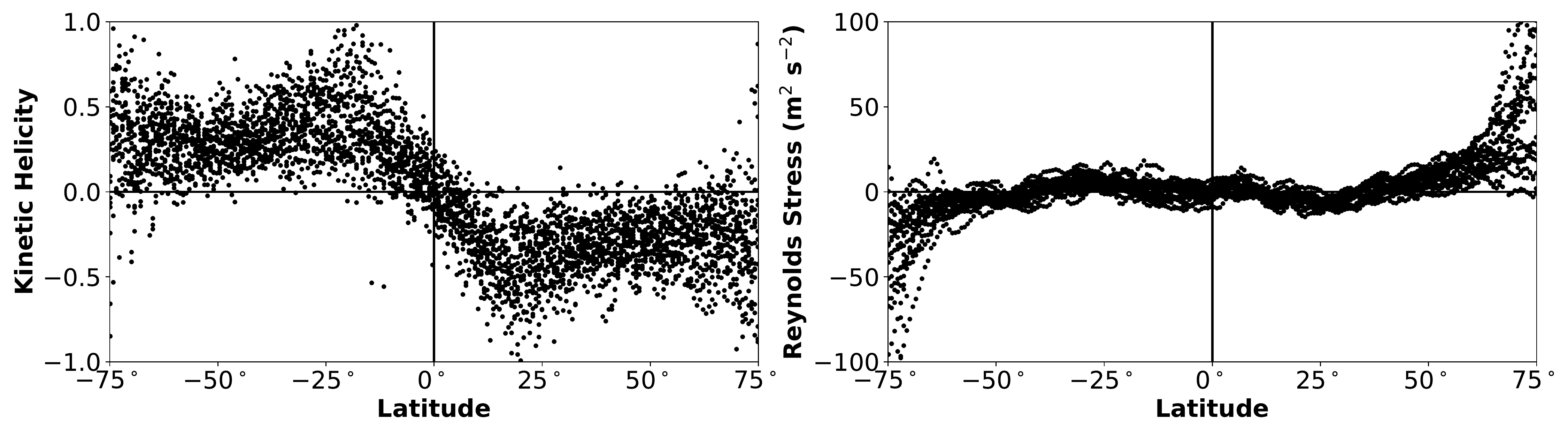}
\caption{Left panel: Kinetic helicity $<\nabla\cdot{V_H} ~\nabla\times{V_H}>$ of the giant cells measured over 19 6-month intervals. The kinetic helicity is negative (left-handed) in the north and positive (right-handed) in the south but with very different latitudinal variations at low vs. high latitudes. Right panel: Reynolds stress $<V_\theta V_\phi>$ of the giant cells measured over 19 6-month intervals. The angular momentum flux is near zero at the low latitudes but strongly toward the equator at high latitudes.}
\label{fig:GCKHRS}
\end{figure}

The effects of the Sun's rotation on these large, long-lived cellular flows is expected to produce key signatures in both the kinetic helicity (the dot product of the velocity with the vorticity) and the component of the Reynolds stress given by the (anti) correlation of poleward flows and prograde flows ($<V_\theta V_\phi>$).

We produce a proxy, $<\nabla\cdot{V_H} ~\nabla\times{V_H}>$, for the radial component of the kinetic helicity by using the divergence of the horizontal flows as an estimate of the relative amplitude and direction of the radial flow vector at each location.
(This association follows from the mass continuity equation.)
The profiles of this kinetic helicity proxy are shown in the left panel of Figure~\ref{fig:GCKHRS}.
The kinetic helicity is negative (left-handed) in the north and positive (right-handed) in the south.
The low- and high-latitude flow regimes are also reflected in this quantity.
The kinetic helicity increases linearly across the equator from about $25^\circ$ north to $25^\circ$ south.
At higher latitudes it is relatively constant (albeit with a slight dip at about $50^\circ$ latitude in each hemisphere).

The Reynolds stress $<V_\theta V_\phi>$ is associated with the latitudinal transport of angular momentum and, as such, it is key to producing and maintaining the Sun's differential rotation with its rapidly rotating equator.
The latitudinal profiles of this component of the Reynolds stress are shown in the right panel of Figure~\ref{fig:GCKHRS}.
The Reynolds stress is small (consistent with zero) at low latitudes.
The Reynolds stress rises rapidly at high latitude to give a strong flux of angular momentum toward the equator.

\section{Spherical Harmonic Spectra} \label{sec:Spectra}

We produced spherical harmonic spectra of the giant cell flows by first constructing maps of the curl and of the divergence of the horizontal velocity.
This provides us with measures of the two spherical harmonic components - the toroidal and poloidal components.

Following \cite{Chandrasekhar61} we can fully represent the horizontal flows on the surface of the Sun using toroidal, $T_\ell^m$, and poloidal, $S_\ell^m$, spectral coefficients with

\begin{equation}\label{eqn:V_theta}
    V_\theta(\theta,\phi) = \sum_\ell \sum_{m=-\ell}^\ell \lbrack
    S_\ell^m {\partial Y_\ell^m \over \partial\theta}
    + T_\ell^m {1 \over \sin\theta} {\partial Y_\ell^m \over \partial\phi} \rbrack
\end{equation}

\begin{equation}\label{eqn:V_phi}
    V_\phi(\theta,\phi) = \sum_\ell \sum_{m=-\ell}^\ell \lbrack
    S_\ell^m {1 \over \sin\theta} {\partial Y_\ell^m \over \partial\phi}
    - T_\ell^m {\partial Y_\ell^m \over \partial\theta} \rbrack
\end{equation}

\noindent
where

\begin{equation}
Y_\ell^m(\theta,\phi) = \bar{P}_\ell^m(\cos\theta) e^{im\phi}
\end{equation}

\noindent
is a normalized spherical harmonic of degree $\ell$ and order $m$.

By choosing this description of the horizontal velocities, the divergence of the flow field gives

\begin{equation}
\nabla\cdot{V_H} = {1 \over r} \sum_\ell \sum_{m=-\ell}^\ell \ell (\ell + 1) S_\ell^m Y_\ell^m(\theta,\phi)
\end{equation}

\noindent
and the curl of the flow field gives

\begin{equation}
\nabla\times{V_H} = - {1 \over r} \sum_\ell \sum_{m=-\ell}^\ell \ell (\ell + 1) T_\ell^m Y_\ell^m(\theta,\phi)
\end{equation}

The mean squared velocity on the surface is given by

\begin{equation}
V_{RMS}^2 = {1 \over 4 \pi} \int_0^\pi \int_0^{2\pi} [V_\theta^2 + V_\phi^2] d\phi \sin\theta d\theta
\end{equation}

\noindent
which becomes

\begin{equation}
V_{RMS}^2 = \sum_\ell \sum_{m=-\ell}^\ell \ell (\ell + 1) [{S_\ell^m}^2 + {T_\ell^m}^2]
\end{equation}

With this description in mind, we find the spherical harmonic spectral coefficients for the giant cellular flows by first taking the divergence of the giant cell flow velocities (Figure~\ref{fig:GCFlowMaps}) to isolate the poloidal components and by taking the curl to isolate the toroidal components.
Performing a spherical harmonic transform on the divergence and curl maps then gives the individual spectral coefficients.

\begin{figure}[htb]
\includegraphics[width=1.0\columnwidth]{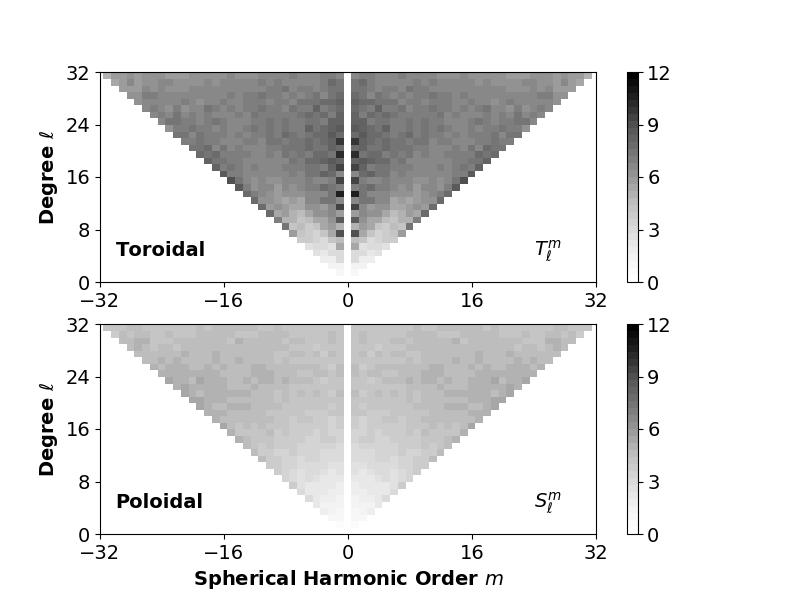}
\caption{Spectral amplitudes as functions of spherical harmonic degree $\ell$ and azimuthal order $m$. The averages of the toroidal (vorticity) coefficients are shown in the upper panel. The averages of the poloidal (divergence) coefficients are shown in the lower panel. The toroidal coefficients dominate and show signatures of distinct high- and low-latitude components at $m=\pm1$ and $m=\pm\ell$ respectively.}
\label{fig:Spectrum2D}
\end{figure}

The average spectral amplitudes are shown in Figure~\ref{fig:Spectrum2D} with the toroidal components in the top panel and the poloidal components in the bottom panel.
The toroidal coefficients are clearly much stronger than the poloidal coefficients - indicating that the flows are dominated by vortices with significant curl and little divergence.
(Note that, at each latitude, the longitudinal average of each component of the giant cellular flows was removed from each maps since these represent axisymmetric flows -- hence the lack of any power at $m=0$ in Figure~\ref{fig:Spectrum2D}.)

Figure~\ref{fig:Spectrum2D} also shows signatures for two different components within the toroidal coefficients - a strong component at $m=\pm1$ (the high latitude component) and a weaker but clearly evident component with $m=\pm\ell$ (the equatorial component).

\begin{figure}[htb]
\includegraphics[width=1.0\columnwidth]{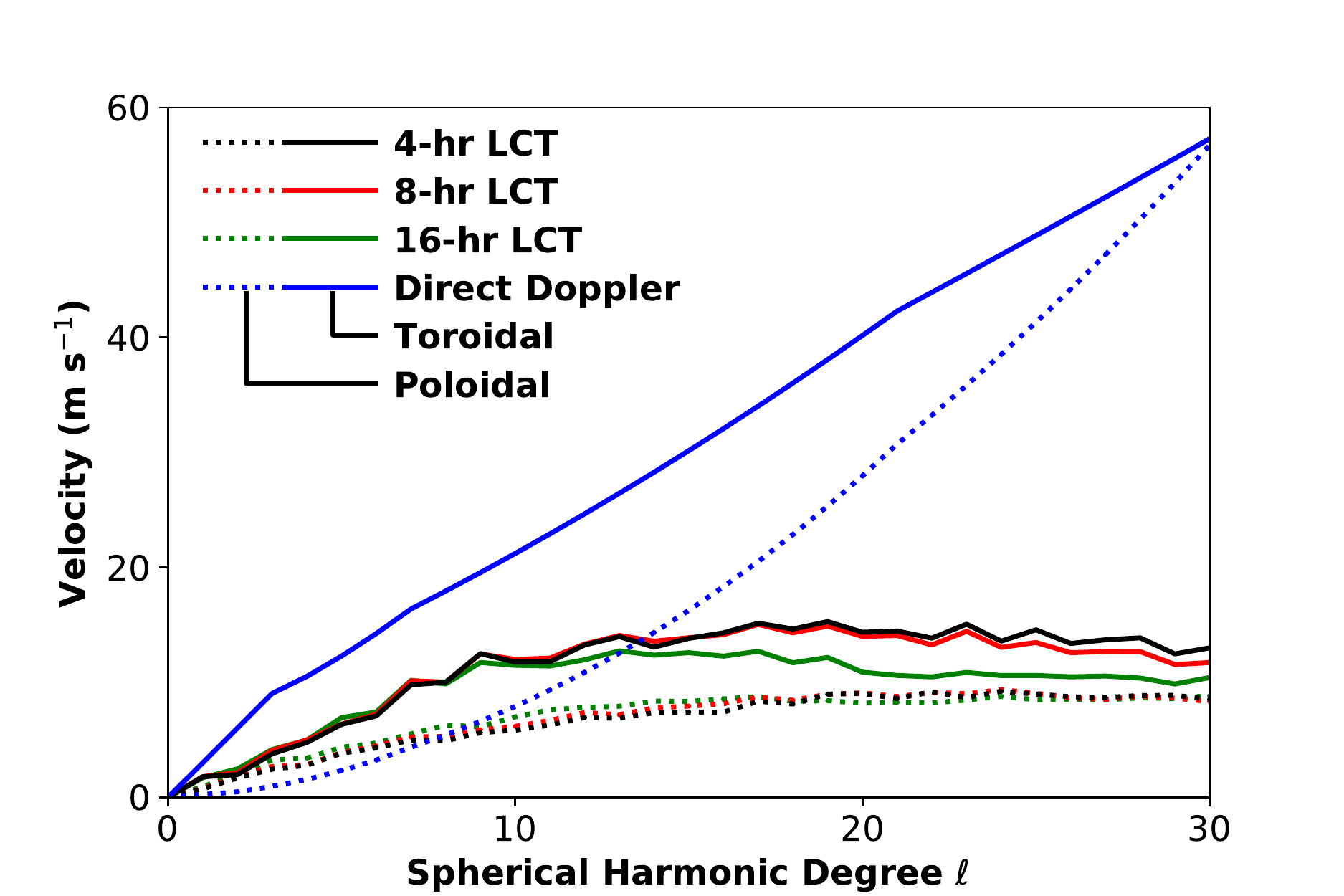}
\caption{Characteristic velocities as functions of spherical harmonic degree $\ell$. The toroidal components (solid lines) dominate the poloidal components (dotted lines) in this wavenumber range. Spectra derived from measurements with 4-hr time lags are shown in black, those derived from measurements with 8-hr time lags are shown in red, and those derived from measurements with 16-hr time lags are shown in green. Spectra derived from direct Doppler measurements by \cite{Hathaway_etal15} are shown in blue. The characteristic velocities increase monotonically -- the turnover at wavenumbers above $\ell\sim20$ is due to the drop-off in sensitivity associated with the 90 Mm window diameter used in the LCT step. The direct Doppler measurements overestimate the spectral amplitudes at these wavenumbers.}
\label{fig:Spectrum1D}
\end{figure}

The characteristic (RMS) velocity associated with each component is shown as a function of spherical harmonic degree $\ell$ in Figure~\ref{fig:Spectrum1D}.
The toroidal component dominates -- as expected from the effects of the Sun's rotation.
(Note the characteristic velocity of $\sim12$ m s$^{-1}$ at $\ell=10$.)
While the characteristic velocities seem to fall-off at wavenumbers above $\ell\sim20$, this is attributed to the drop in sensitivity at higher wavenumbers.
While we make LCT measurements at 256 locations in latitude and longitude from each pair of Doppler maps, the 21-pixel (90 Mm) diameter correlation window used in the LCT step only gives 24 fully independent samples in latitude.
Figure~\ref{fig:Spectrum1D} also includes, in red, the spectra from giant cell flow maps constructed from LCT measurements made with an 8-hr time lag and, in green, the spectra from giant cell flow maps constructed from LCT measurements made with a 16-hr time lag.
These measurements represent deeper layers (22 Mm at 4-hr, 26 Mm at 8-hr, and 37 Mm at 16-hr time lags according to \cite{Hathaway12B}) within the surface shear layer but still exhibit virtually the same velocity spectra for the giant cell flows in those deeper layers.

The velocity spectra derived from direct Doppler measurements in \cite{Hathaway_etal15} are shown in blue in Figure~\ref{fig:Spectrum1D}.
While these spectral are not subject to any decrease in amplitude due to loss of resolution at these wavenumbers, they are subject to contamination due to instrumental artifacts in the Doppler data itself (see discussion in \cite{Hathaway_etal15}).
The LCT measurements are somewhat less than the direct Doppler measurements at the well resolved wavenumbers (below $\sim10$).
We expect that the low wavenumber LCT spectral amplitudes are better representative of the actual flows on the Sun and that the direct Doppler measurements are contaminated by instrumental artifacts at these wavenumbers.
It is also important to note that the direct Doppler spectra indicate equal parts toroidal and poloidal flow at $\ell=30$ while the LCT measurements indicate that this cross-over must occur at a higher wavenumber.

\section{Equatorial Rossby Waves} \label{sec:RossbyWaves}

Wave-like phenomena are characterized by taking the Fourier transform in time of the individual spectral coefficients shown in Figure~\ref{fig:Spectrum2D}.
Our giant cell flow maps and spherical harmonic spectra of the vorticity are generated at a cadence of 1 per day.
We examine the temporal variation by taking the Fourier transform in time of 1024-day records sampled at 6-month intervals (14 such overlapping records are contained in the SDO/HMI mission to date).
A cosine taper (of total length 256) was applied to each end of each spectral coefficient time series prior to taking the Fourier transform to reduce ringing due to end effects.

\begin{figure}[htb]
\includegraphics[width=1.0\columnwidth]{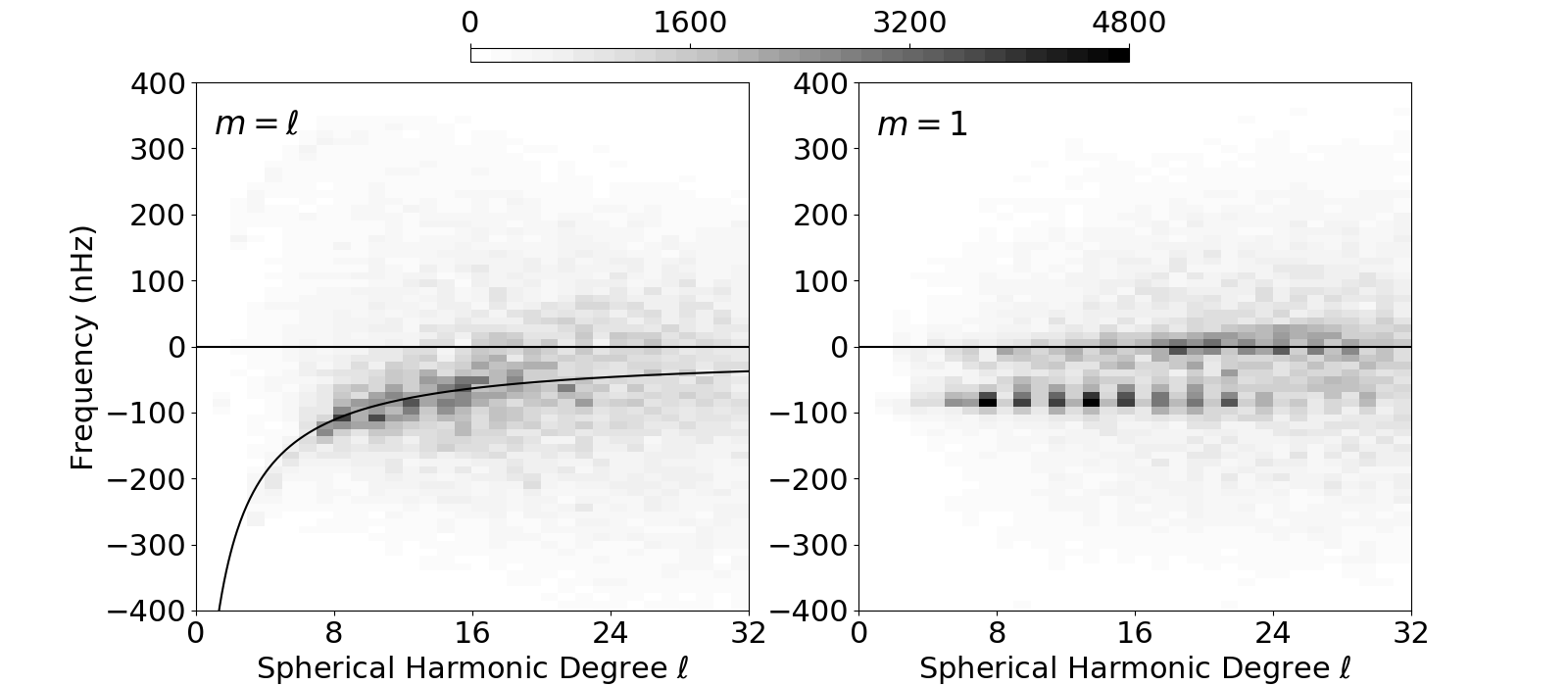}
\caption{Vorticity spectral amplitudes as functions of spherical harmonic degree $\ell$ and temporal frequency for modes with $m=\ell$ (left) and with $m=1$ (right). The functional form of the dispersion relation for Rossby waves given by Eq.~\ref{eqn:k_omega} is shown as the curved line at negative frequencies on the left.}
\label{fig:GCk_omega}
\end{figure}

The toroidal spectral amplitudes as functions of temporal frequency and wavenumber, $\ell$, for modes with $m=\ell$ (equatorial modes) are shown in the left panel in  Figure~\ref{fig:GCk_omega}.
These modes propagate retrograde with a dependence upon wavenumber very much like the dispersion relation given for Rossby waves or r-modes \citep{Saio82}

\begin{equation}\label{eqn:k_omega}
\omega = {-2\Omega_\odot m \over [\ell(\ell + 1)]}
\end{equation}

\noindent
where $\Omega_\odot/2\pi = 456$ nHz is the Carrington rotation frequency.
This functional form is indicated by the curved line in the left panel of Figure~\ref{fig:GCk_omega}.
This Rossby wave characteristic of the equatorial modes was first noted by \cite{Loptien_etal18}.

Rossby waves, as derived by \cite{Rossby39}, are waves that arise in shallow water or thin atmospheric layers when the characteristic time scale for the flows are much longer than the rotation period.
Under these circumstances the flows become toroidal (geostrophic) with the Coriolis force acting on the flows balanced by the horizontal pressure gradients associated with the wave disturbance.
They gain their propagation characteristics from the latitudinal variation in strength of the radial component of the rotation vector -- the component that produces the Coriolis on the horizontal flows.

R-modes are extensions of these waves to deep spherical shells -- initially the stably stratified radiative zones of massive early-type stars \citep{Saio82} .
\cite{WolffBlizard86} explored their properties in the Sun in the absence of convective motions.
Models of solar convection typically find that the convective structures propagate prograde without these r-mode characteristics (see \cite{Miesch05} and references therein). 

\section{Polar Vortices} \label{sec:PolarVortices}

The polar vortices are represented by the toroidal components with $m=\pm1$ in Figure~\ref{fig:Spectrum2D}.
(Toroidal components with $m=\pm2$ and $m=\pm3$, while weaker, also contribute to the structure of the polar vortices.)
The spectral amplitudes as functions of temporal frequency and wavenumber, $\ell$, for modes with $m=\pm1$ are shown in the right panel of Figure~\ref{fig:GCk_omega}.
These modes propagate retrograde with no variation in temporal frequency with wavenumber -- indicating that these high latitude modes are advected retrograde by the differential rotation at a rate about 80 nHz slower than the 456 nHz Carrington reference rate.
This matches the rotation rate at $\sim 60^\circ$ latitude as seen in Figure~\ref{fig:GCDRMF}.
The same behavior is also seen for the $m=\pm2$ and $m=\pm3$ components but with nearly double and triple the temporal frequencies, giving similar rotation rates.
Note that the $m=\pm1$ components are dominated by the odd $\ell$ modes (symmetric across the equator) while the $m=\pm3$ components are dominated by the even $\ell$ modes (anti-symmetric across the equator).
The $m=\pm2$ components do not show any preference for symmetry across the equator.

\section{Conclusions} \label{sec:Conclusions}

\begin{figure}[htb]
\includegraphics[width=1.0\columnwidth]{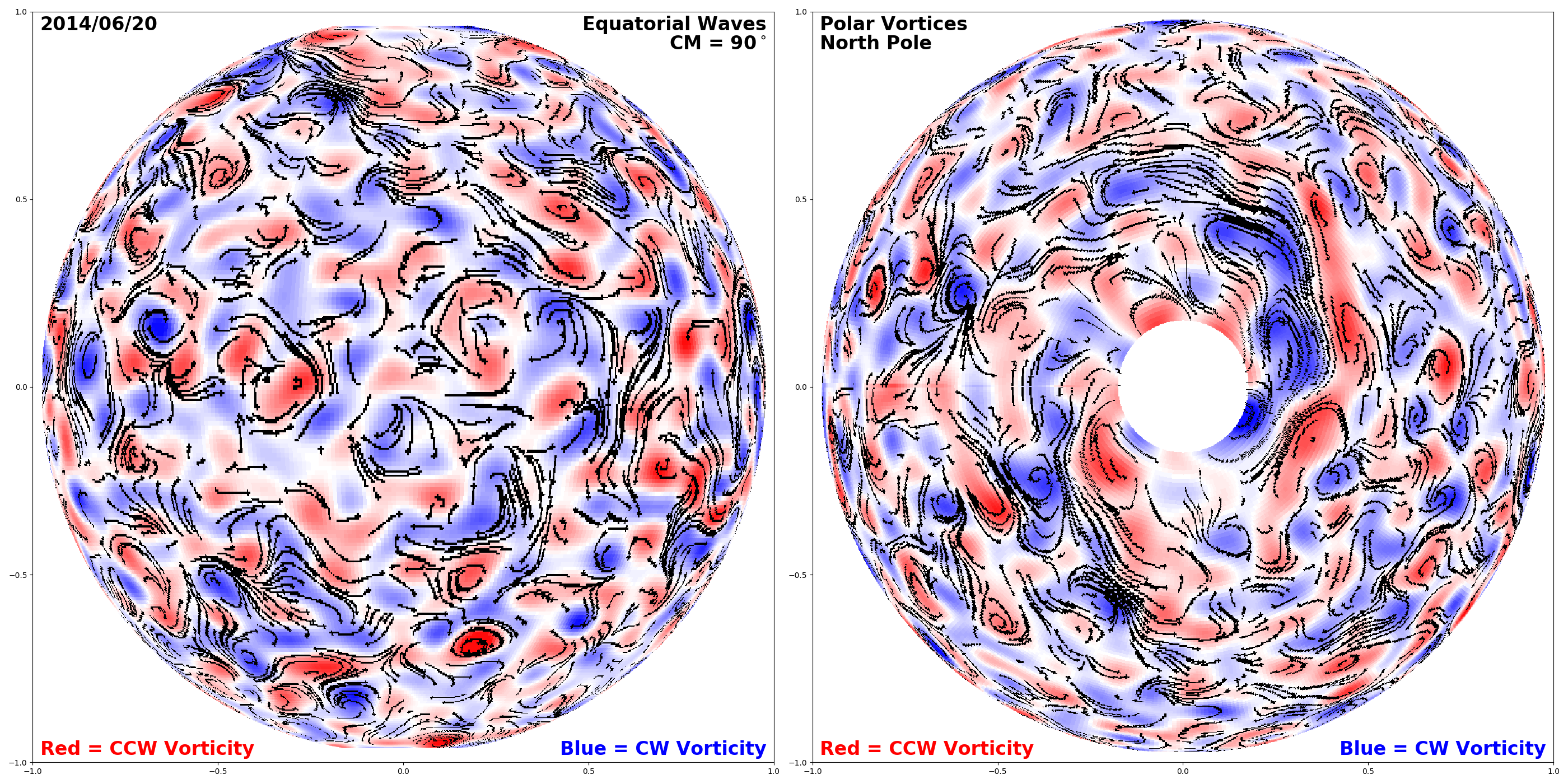}
\caption{Giant cellular flow patterns as seen from above the equator (left) and from above the north pole (right). The underlying colored image shows the vorticity (right-handed in red and left-handed in blue). Flow tracers are shown in black with streaklines indicating flow direction. The associated animation is \href{http://solarcyclescience.com/bin/SolarVortices.mp4}{SolarVortices.mp4}}
\label{fig:GiantCells}
\end{figure}

We have improved upon the LCT method used in \cite{Hathaway_etal13} and, by including information on the differential rotation of the giant cellular flows, improved our maps of the giant cellular flows.
We find strong evidence, from multiple characteristics, for the existence of two very different flow regimes -- low latitudes Rossby waves and high latitude polar vortices.
These flows are illustrated in Figure~\ref{fig:GiantCells} and the associated animation \href{http://solarcyclescience.com/bin/SolarVortices.mp4}{SolarVortices.mp4}.

The low latitude Rossby waves are nearly circular in shape as indicated by visual inspection of the giant cell flow maps (Figure~\ref{fig:GCFlowMaps}) and by the auto-correlation study illustrated in Figure~\ref{fig:GCAutoCor}.
They are not extended north to south as suggested by nearly every numerical model of solar convection zone dynamics (cf. \cite{Miesch05} and references therin) and by reports of north/south alignment of supergranules \citep{Lisle_etal04}.
They are also not extended east to west as reported by \cite{Beck_etal98}.

These low latitude Rossby waves have lifetimes only slightly longer than the 27 day rotation period of the Sun.
This may be due the waves getting in and out of phase with each other as the low wavenumber waves propagate faster than the higher wavenumber waves (Eq.~\ref{eqn:k_omega}).

The low latitude Rossby waves do exhibit the kinetic helicity signatures associated with the effects of the Sun's rotation (Figure~\ref{fig:GCKHRS}) with left-handed helicity in the north and right-handed helicity in the south.
They do not, however, exhibit any Reynolds stress like that needed to maintain the Sun's differential rotation.

The low latitude Rossby waves propagate retrograde relative to the Carrington rotation frame of reference, do not exhibit substantial differential rotation with latitude, and do not propagate either poleward or equatorward (Figure~\ref{fig:GCDRMF}).
Spectral analyses of these low latitude flows show clear evidence for a dispersion relation like that found for Rossby waves \citep{Rossby39} or r-modes \citep{Saio82}.
This constitutes further confirmation of the discoveries of these waves by \cite{Loptien_etal18} using LCT of granules and by \cite{Proxauf_etal20} using helioseismology.
Our observations extend the measurements of these waves to greater depths.
The granules tracked by \cite{Loptien_etal18} extend to depths of $\sim1$ Mm while the acoustic waves used by \cite{Proxauf_etal20} reliably sample flows at depths down to $\sim8$ Mm.
Our LCT measurements of supergranules at 4-hr, 8-hr, and 16-hr time lags represent flows at depths of 22 Mm, 26 Mm, and 37 Mm respectively.
While \cite{Proxauf_etal20} conclude that the Rossby wave amplitudes decrease by 10\% with depth over the outermost 8 Mm, we find that the spectral amplitudes are constant down to far greater depths.

The high latitude polar vortices differ morphologically and dynamically from the low latitude Rossby waves.
The polar vortices are long, narrow features that spiral into the polar regions as indicated by visual inspection of the giant cell flow maps (Figure~\ref{fig:GCFlowMaps}), by the auto-correlation study (Figure~\ref{fig:GCAutoCor}), and by the flow structures shown in Figure~\ref{fig:GiantCells} and the associated animation (\href{http://solarcyclescience.com/bin/SolarVortices.mp4}{SolarVortices.mp4}).

The polar vortices have lifetimes of several months.
These features are not well represented in numerical models of the solar convection zone but they have been seen via measurements with other methods.
\cite{Bogart_etal15} used the helioseismic method of ring diagram analysis and found high latitude features that match those found here in terms of shape, position, orientation, and lifetimes -- but with far weaker flow velocities (only 0.5 m s$^{-1}$ instead of 12  m s$^{-1}$).
\cite{Howe_etal15} also used ring diagram analysis and, while they didn't show maps of the velocity structures, they did find velocity structures with velocities more commensurate with ours ($\sim20$ m s$^{-1}$).

While \cite{Hathaway_etal13} concluded that the flow velocities in these giant cellular flows increased with depth, we now find that the velocity spectra are unchanged with depth (Figure~\ref{fig:Spectrum1D}).
The previous conclusion was based on increased velocities seen in the giant cell flow maps themselves and was influenced by the increase in velocity measurement noise at longer time lags.
The velocity spectral amplitudes shown in Figure~\ref{fig:Spectrum1D} are somewhat smaller than those found by \cite{Hathaway_etal15} and they indicate that the cross-over from toroidal flows at low $\ell$ to poloidal flows at high $\ell$ must occur at a wavenumber substantial higher than $\ell=30$ as given by \cite{Hathaway_etal15}.
The LCT measurements employed here are expected to be more reliable than the direct Doppler measurements of \cite{Hathaway_etal15} at these low wavenumbers due to contamination by instrumental artifacts in the direct Doppler data.

The polar vortices exhibit the kinetic helicity signatures associated with the effects of the Sun's rotation (Figure~\ref{fig:GCKHRS}) with left-handed helicity in the north and right-handed helicity in the south.
Unlike the low latitude Rossby waves, they exhibit a strong Reynolds stress giving an equatorward transport of angular momentum needed to help maintain the Sun's differential rotation.
This Reynolds stress can be attributed to the shape and orientation of the features themselves.
The Sun's rotation forces the flows themselves to be directed along the structures and the structures are oriented in a spiral pattern.

The polar vortices rotate differentially with 34 day periods above $75^\circ$ latitude, 30 days at $60^\circ$ latitude, and 28 days at $45^\circ$ latitude (Figure~\ref{fig:GCDRMF}).
This differential rotation profile differs from that seen near the surface (the magnetic element rotation profile shown in red in Figure~\ref{fig:GCDRMF} is representative of the surface shear layer at a depth of about 20 Mm).
The faster rotation rates of the polar vortices (relative to the surface) are much more representative of the rotation near the base of the convection zone (cf. \cite{Howe09}).
The rotation rates of the polar vortices at $60^\circ$  and $75^\circ$ match the internal rotation rates at those latitudes at one and only one radius, $\sim0.78$ R$_\odot$ -- the top of the tachocline at the base of the convection zone.

The polar vortices also exhibit a meridional flow -- poleward at a peak velocity of about 2 m s$^{-1}$.
If these vortices do extend to the base of the convection zone where we find similar rotation rates, then this meridional flow observation indicates a slow poleward meridional flow at the same depth.

The radial (and latitudinal) structure of the Sun's meridional circulation has very important consequences for models of the Sun's magnetic dynamo.
Flux transport dynamos \citep{Choudhuri_etal95, NandyChoudhuri02} assume that the poleward meridional flow seen near the surface sinks inward in the polar regions and returns equatorward at the base of the convection zone where they expect the sunspot magnetic fields to arise - thus giving rise to the equatorward drift of the sunspot zones.
However, recent observations of an equatorward meridional flow at the base of the surface shear layer\citep{Hathaway12B, Zhao_etal13} and now this observation suggesting poleward flow at the base of the convection zone (in agreement with the double cell meridional structure suggested by the observations of \cite{Zhao_etal13}) further challenge these Flux Transport Dynamo models.

The hydrodynamic properties of these giant cellular flows are not well represented (if at all) in current numerical models of the Sun's convection zone dynamics (neither the high latitude spirals nor the low latitude Rossby waves are found in these models).
This suggests that these models (which undoubtedly solve the proper hydrodynamic and MHD equations) might require further development of surface boundary conditions and/or initial conditions.
It is well worth noting that the Sun previously rotated faster and had a lower luminosity - conditions that may have led to the solar differential rotation with flows like these polar vortices that continue to this day to maintain the solar differential rotation as a positive feedback mechanism.
Another possibility is that the low latitude Rossby waves may be a rather shallow phenomena that hides underlying structures more representative in models simulations at low latitudes.

\acknowledgments

The HMI data used here are courtesy of the NASA/SDO and the HMI science teams.
The authors benefited from discussions with Sushant Mahajan on improving the LCT method and from comments on the manuscript by Phil Scherrer, Todd Hokesema, Leif Svalgaard, and Adam Hathaway.
DHH was supported by the HMI science team at Stanford University.
L.A.U. was supported by the NSF Atmospheric and Geospace Sciences Postdoctoral Research Fellowship Program (Award Number:1624438) and the NASA Living with a Star Program (Grants:NNH16ZDA001N-LWS and NNH18ZDA001N-LWS). 

\bibliography{Hathaway}{}
\bibliographystyle{aasjournal}

\end{document}